\documentclass[aps, prl, reprint, superscriptaddress, floatfix,]{revtex4-1}
\usepackage{amsmath, amssymb, graphicx}
\usepackage[hidelinks]{hyperref}

\begin{document}

\title{Symmetry-protected topological magnons in three dimensional Kitaev materials}

\author{Wonjune Choi}
\affiliation{Department of Physics, University of Toronto, Toronto,
  Ontario M5S 1A7, Canada}
\author{Tomonari Mizoguchi}
\affiliation{Department of Physics, University of Tsukuba, Tsukuba, Ibaraki 305-8577, Japan}
\author{Yong Baek Kim}
\affiliation{Department of Physics, University of Toronto, Toronto,
  Ontario M5S 1A7, Canada}
\affiliation{Canadian Institute for Advanced Research, Toronto, Ontario M5G 1Z8, Canada}

\begin{abstract}
Topological phases in magnetic materials offer novel tunability of topological properties via varying  
the underlying magnetism. We show that three dimensional Kitaev materials can provide a great 
opportunity for controlling symmetry-protected topological nodal magnons. These materials are originally
considered as strong candidates for the Kitaev quantum spin liquid due to the bond-dependent
frustrating spin exchange interactions. As a concrete example, we consider the symmetry and topology
of the magnons in the canted zig-zag ordered state in the hyperhoneycomb $\beta$-Li$_2$IrO$_3$,
which can be obtained by applying a magnetic field in the counter-rotating spiral state at zero field.
It is shown that the magnetic glide symmetries and the non-Hermitian nature of the bosonic magnons
lead to unique topological protection that is different from the case of the fermionic counterparts. 
We investigate how such topological magnons can be controlled by changing the symmetry of 
the underlying spin exchange interactions.
\end{abstract}

\maketitle

Transition metal oxides/halides with strong spin-orbit coupling 
on trivalent lattices are getting much attention in the past few years \cite{William_review, Rau_review, Winter_2017,Trebst_2017,Schaffer_2016}, due to 
the possibility of realizing the celebrated Kitaev spin liquid, a quantum disordered state 
whose spins are fractionalized into Majorana fermions \cite{Kitaev_Honeycomb, JKmechanism, Kimchi_3dIridate,Chaloupka_JKmodel,JKhyperhoneycomb,Takayama_hyperhoneycomb,Katukuri,U1hyperhoneycomb}.
Most of these Kitaev materials are magnetically ordered at ambient pressure 
and zero magnetic field \cite{hyperhoneycomb_spiralExperiment, harmonicExp, 3dharmonicHoneycomb}, due to the presence of other anisotropic exchange interactions 
in addition to the Kitaev's bond-dependent Ising interaction between 
the spin-orbit-coupled $j_\mathrm{eff}=1/2$ moments \cite{Lee_hyperhoneycomb, Rau_JKGmodel, Chaloupka_JKmodel, JKhyperhoneycomb, Lee_twoIridate}.
Recent efforts have focused on the suppression of the magnetic order
to achieve the putative spin liquid state \cite{Kasahara2018, Takagi_hyperhoneycomb, Banerjee2018, Baek2017, Yadav2016, Balz2019, Veiga_hyperhoneycomb, Clancy2018, Kim_RuClBfield}.
So far limited effort has been 
made to investigate possible topological phenomena in the magnetically
ordered states in these systems \cite{McClarty_KitaevMagnon, Lu_2018}.

\begin{figure}[t]
\centering
\includegraphics[width=0.5\textwidth]{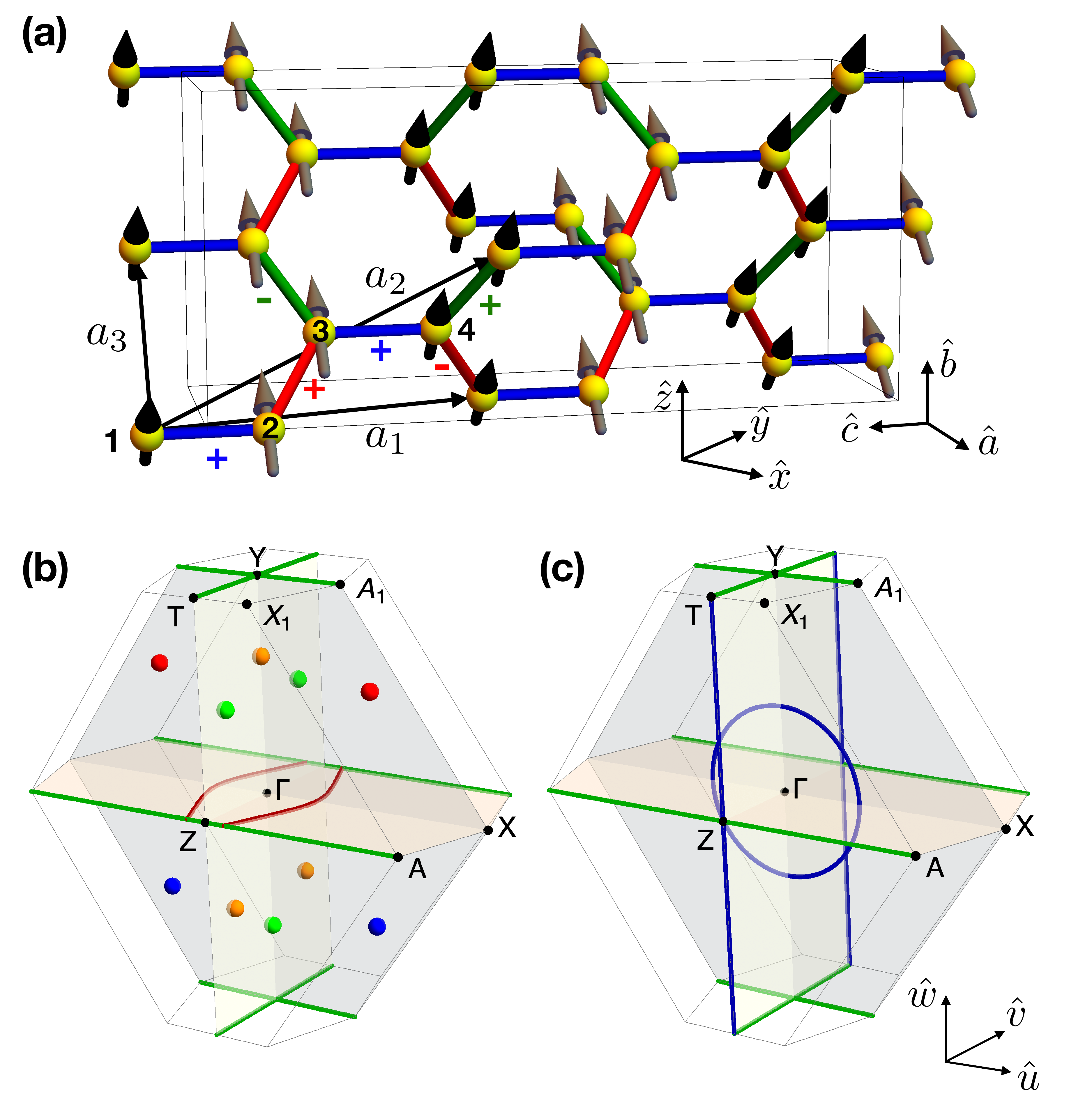}
\caption{(a) Field-induced canted zig-zag order. The sign $\xi_{jk}$ for the off-diagonal exchange is shown.
(b) Topological magnons with $(J, K, \Gamma, h) = (0.063, -1, -0.33, 0.065)$. The magnetic glides ($\mathcal{T}d_2$, $\mathcal{T}d_3$) protect the green nodal lines, and the glide mirror $d_1$ protects the red nodal lines.
Red (orange) and blue (green) points represent the Weyl magnons between the first (second) and second (third) bands with chirality $-1$ and $+1$, respectively.
(c) Generic magnon band crossing without magnon pairing.
The magnon pairing gaps out the blue nodal lines.
}
\label{fig:magnon}
\end{figure}

In this paper, we theoretically study the symmetry-protected topological
nodal magnons in three dimensional Kitaev materials, whose properties can
be tuned by changing the underlying magnetic order.
It has been known that the novel bond-dependent exchange interactions
lead to the highly unusual counter-rotating incommensurate spiral order
in $\beta\text{-}\mathrm{Li_2IrO_3}$ \cite{hyperhoneycomb_spiralExperiment, Lee_hyperhoneycomb, Ducatman_hyperhoneycomb, Peter_hyperhoneycomb}, where the spin-orbit coupled 
$j_{\rm eff}=1/2$ local moments of Ir$^{4+}$ are located on the three-dimensional
hyperhoneycomb lattice.
Upon the application of an external magnetic field
along the $\hat{b}$-axis, a recent experiment has found the phase transition to a canted zig-zag order [Fig.~\ref{fig:magnon} (a)] \cite{Ioannis_fieldHyperhoneycomb, fieldHyperhoneycomb_experiment}.
As follows, we investigate the role of non-symmorhic magnetic space group symmetries on the topological nature of the  
magnons in this field-induced canted zig-zag order in $\beta\text{-}\mathrm{Li_2 Ir O_3}$.
Even though the magnetic order itself is topologically trivial, its magnon excitations exhibit topologically protected zero and 
one-dimensional gapless band touching dispersions (Fig.~\ref{fig:magnon}).
By constructing the symmetry-constrained generic magnon Hamiltonian, we explicitly show that the magnon pairing 
originating from the spin-orbit coupling is responsible for the glide-protected nodal lines and Weyl points.
The magnon pairing and the non-conservation of the magnon number reflect the non-Hermitian nature
of bosonic magnon systems.
Unlike fermionic counterparts, these gapless nodal lines/points are not obtained by partially gapping out higher dimensional gapless band crossings.
Instead, they are born out of the bosonic statistics of magnons, which realizes the magnon spectrum in a non-Hermitian fashion.
Using the magnetic space group symmetry and the classification of non-Hermitian topological phases \cite{non-HermitianTable, nonHermitian2},
we find that the integer topological invariants guarantee the stability of these topological magnon spectra.

\textit{Model.} -- The minimal spin model for $\beta$-$\mathrm{Li_2IrO_3}$, so called $JK\Gamma$ model \cite{Lee_hyperhoneycomb, Rau_JKGmodel}, consists of the Heisenberg interaction ($J$), the Kitaev interaction ($K$), and the off-diagonal exchange ($\Gamma$) between the nearest-neighbor local moments:
\begin{align}
H_{JK\Gamma} &= \sum_{\alpha\textrm{-link}} J \vec{S}_j \cdot \vec{S}_k+ K S_j^\alpha S_k^\alpha + \xi_{jk} \Gamma (S_j^\beta S_k^\gamma + S_j^\gamma S_k^\beta),
\label{eq:JKGmodel}
\end{align}
where $\alpha=x,y,z$ denotes three different types of the nearest-neighbor links, and the lattice symmetry determines the sign $\xi_{jk} = \pm 1$ for the off-diagonal exchange [Fig.~\ref{fig:magnon}~(a)]\cite{Lee_hyperhoneycomb}.
This model not only reproduces the known zero-field ground state, incommensurate non-coplanar spiral \cite{Lee_hyperhoneycomb, Ducatman_hyperhoneycomb, Peter_hyperhoneycomb, hyperhoneycomb_spiralExperiment}, but also the recently discovered field-induced canted zig-zag order \cite{Ioannis_fieldHyperhoneycomb, fieldHyperhoneycomb_experiment} under a sufficiently strong magnetic field along $\hat{b}$-axis ($S^z$ direction), $H_h = -h \sum_j S_j^z$.

Based on first principle calculations \cite{Kim_DFT}, the experimentally relevant set of parameters for $\beta$-$\mathrm{Li_2IrO_3}$ is $(J,K,\Gamma)=(0.063,-1,-0.33)$ in energy unit $|K|=1$ \cite{SM}.
Since the Heisenberg interaction is much smaller than the other two interactions, we focus on the system with varying strength of ferromagnetic $\Gamma < 0$ and fixed $J=0.063$.

\textit{Linear spin wave theory}. --
Given the field-induced magnetic ground state, we study the magnon excitations within linear spin wave theory.
We construct a generic magnon Hamiltonian based on the symmetries of the magnetic order and range of interactions.
While the $JK\Gamma$ model on the hyperhoneycomb lattice respects time-reversal symmetry $\mathcal{T}$ and $Fddd$ space group  generated by three glide mirror planes ($d_1$, $d_2$, $d_3$) (definitions of $d_{1,2,3}$ are in Supplementary Material \cite{SM}), an external magnetic field and induced magnetic order explicitly/spontaneously break time-reversal symmetry $\mathcal{T}$ and two glide mirror symmetries $d_2$ and $d_3$.
However, the magnetic order is still invariant under the product of time-reversal and glide plane symmetries,
\begin{align}
&\mathcal{T}d_2 : (S_j^x, S_j^y, S_j^z) \to \left(S_{d_2(j)}^y, S_{d_2(j)}^x, S_{d_2(j)}^z\right), \\
&\mathcal{T}d_3 : (S_j^x, S_j^y, S_j^z) \to \left(-S_{d_3(j)}^y, -S_{d_3(j)}^x, S_{d_3(j)}^z\right).
\end{align}
Therefore, the symmetry group for the canted zig-zag order is the magnetic space group $Fdd'd'$ generated by one glide mirror plane $d_1$ and two \emph{magnetic glides} $d'_2 = \mathcal{T} d_2$ and $d'_3=\mathcal{T}d_3$.

If the physical Hamiltonian only possesses the nearest-neighbor interactions, the $Fdd'd'$ symmetry-constrained generic magnon Hamiltonian is of the form
\begin{align}
\hat{H} = \frac{1}{2} \sum_{\mathbf{q},l,l'}
\begin{pmatrix}
b_{\mathbf{q}l}^\dagger & b_{\mathbf{-q},l}
\end{pmatrix}
\left(\mathcal{H}_\mathrm{hop} + \mathcal{H}_\mathrm{pair} \right)
\begin{pmatrix}
b_{\mathbf{q}l'} \\
b_{\mathbf{-q},l'}^\dagger
\end{pmatrix}
\end{align}
with
\begin{align}
&\mathcal{H}_\mathrm{hop} = a_0 + b_0 \sigma^1 \nonumber \\
&+x_0\left( A_{11} s^1 \sigma^1 + A_{21} s^2 \sigma^1+ A_{12} s^1 \sigma^2 + A_{22} s^2 \sigma^2 \right) \label{eq:HSU2},
\end{align}
\begin{align}
&\mathcal{H}_\mathrm{pair} = b_1 \sigma^1 \tau^1 \nonumber \\
&+x_1\left( A_{11} s^1 \sigma^1 + A_{21} s^2 \sigma^1 + A_{12} s^1 \sigma^2 + A_{22} s^2 \sigma^2 \right) \tau^1 \nonumber \\
&+ y_2 \left( B_{11} s^1 \sigma^1 + B_{21} s^2 \sigma^1 + B_{12} s^1 \sigma^2 + B_{22} s^2 \sigma^2\right) \tau^2,
\label{eq:Hpair}
\end{align}
where $s^\alpha, \sigma^\alpha, \tau^\alpha$ are the Pauli matrices, $A$ and $B$ are momentum dependent coefficients, and the other coefficients are constants \cite{SM}.
We relabel four sublattice sites $l=2s'+\sigma'+1$ with two flavors $s'=0,1$ and $\sigma'=0,1$. $s^\alpha$ and $\sigma^\alpha$ act on these two-dimensional flavor spaces, and $\tau^\alpha$ acts on the particle-hole space.
Since the six parameters $a_0, b_0, b_1, x_0, x_1, y_2$ are implicit functions 
of $J, K, \Gamma$, and $h$, we can deduce their numerical values by constructing the magnon Hamiltonian directly from $JK\Gamma h$ model.

Note that our magnon Hamiltonian has not only the hopping terms $\mathcal{H}_\mathrm{hop}$ but also magnon pairing terms $\mathcal{H}_\mathrm{pair}$.
Due to the bosonic statistics, $[ b_i, b_j^\dagger ] = - [b_j^\dagger, b_i]$, a unitary transformation is no longer canonical in the presence of pairing; the bosonic Bogoliubov-de Gennes (BdG) Hamiltonian should be diagonalized with a paraunitary transformation $T$ such that $T\tau^3 T^\dagger = T^\dagger \tau^3 T  = \tau^3$.
Then the energy spectrum of the \emph{Hermitian} magnon Hamiltonian $\hat{H}$ comprises of the eigenvalues of a \emph{non-Hermitian} effective Hamiltonian $\tau^3 \left( \mathcal{H}_\mathrm{hop} + \mathcal{H}_\mathrm{pair} \right)$ \cite{bosonDiag, Shindou_magnon}.
Therefore, the emergence of topological magnons with pairing is essentially related to the non-Hermitian topological phases \cite{non-HermitianTable, nonHermitian2,bosonicBdG}.

On physical grounds, magnon pairing $\mathcal{H}_\mathrm{pair}$ is a consequence of the non-collinear magnetic order and the bond-directional interactions originating from the spin-orbit coupled nature of $j_\mathrm{eff}=\frac{1}{2}$ moments.
If the magnetic order were collinear and the physical Hamiltonian had $SU(2)$ spin rotation symmetry, the magnetic space group symmetry would have $\widetilde{S}^z$ rotation invariance, which preserves the number of magnons $b_j^\dagger b_j$ and prevents pairing of bosons.

\begin{figure}[t]
\centering
\includegraphics[width=0.5\textwidth]{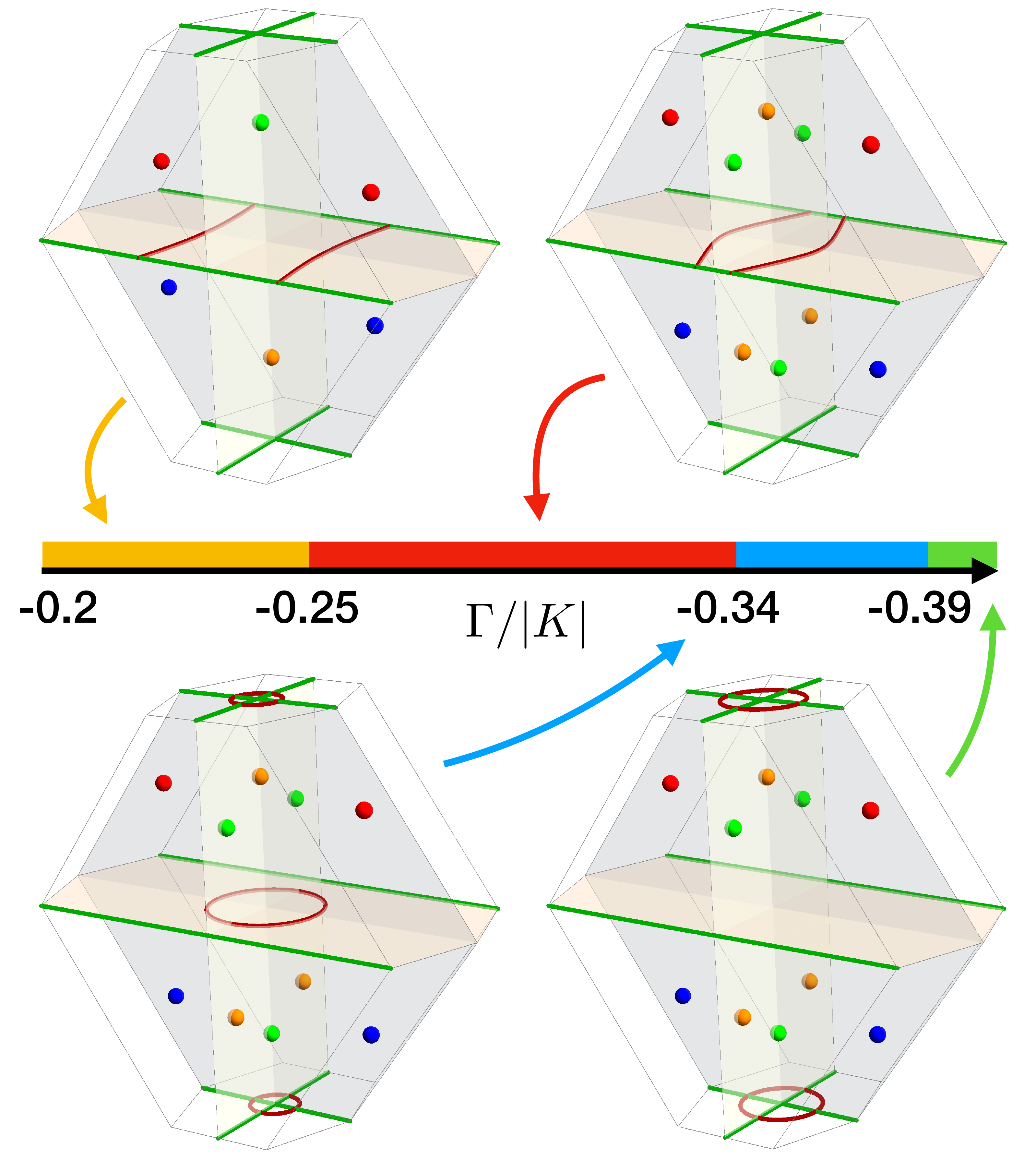}
\caption{Magnon phase diagram with $(J,h)=(0.063,0.065)$.
$\mathcal{T}d_2$ and $\mathcal{T}d_3$ protected green nodal lines always exist.
Strong $\Gamma$ gaps out $d_1$ protected red nodal lines between the first and second bands,
but another $d_1$ protected nodal lines emerge between the third and fourth bands at $\Gamma > -0.34|K|$.
There are four Weyl points between the first and second bands and six Weyl points between the second and third bands at $k_v = 0$ and $k_u=0$ plane, respectively.}
\label{fig:phases}
\end{figure}


\textit{Kramers degeneracy due to magnetic glides.} --
With fixed $J=0.063$ and magnetic field $h=0.065$ above the critical field $h_c$ (pure 
canted zig-zag order is stable for $h > h_c$), we map out the phase diagram for the magnon dispersions (Fig.~\ref{fig:phases}).
Near the experimentally relevant region, $ -0.4 \leq \Gamma \leq -0.2 $, we find two distinct types of nodal lines and a number of Weyl points.

The most robust nodal lines (green lines in Fig.~\ref{fig:phases}) which do not change their locations, originate from the magnetic glide $\mathcal{T}d_2$ and $\mathcal{T} d_3$.
When the magnetic glides act on the momentum state $|q_1, q_2, q_3, l \rangle$, we obtain
\begin{multline}
\mathcal{T}d_2 : \left| q_1, q_2, q_3, l \right \rangle \to \\
\begin{cases}
\left | q_3-q_2,-q_2, q_1 - q_2, 3 \right \rangle^*, & l = 1 \\
\left | q_3-q_2,-q_2, q_1 - q_2,4\right \rangle^*, & l = 2 \\
e^{2\pi i q_2 }\left | q_3-q_2,-q_2, q_1 - q_2,1 \right \rangle^*, & l = 3 \\
e^{2\pi i q_2} \left |q_3-q_2,-q_2, q_1 - q_2,2 \right \rangle^*, & l = 4,
\end{cases}
\label{eq:Td2q}
\end{multline}
\begin{multline}
\mathcal{T}d_3 : \left| q_1, q_2, q_3, l \right \rangle \to \\
\begin{cases}
\left | q_2- q_3,q_1 - q_3, -q_3, 4 \right \rangle^*, & l = 1 \\
\left | q_2- q_3,q_1 - q_3, -q_3,3 \right \rangle^*, & l = 2 \\
e^{2\pi i q_3 }\left | q_2- q_3,q_1 - q_3, -q_3,2 \right \rangle^*, & l = 3 \\
e^{2\pi i q_3} \left | q_2- q_3,q_1 - q_3, -q_3,1 \right \rangle^*, & l = 4,
\end{cases}
\label{eq:Td3q}
\end{multline}
where $\mathbf{q} = q_1 \mathbf{b}_1 + q_2 \mathbf{b}_2 + q_3 \mathbf{b}_3$ and $\mathbf{b}_j$ are the reciprocal lattice vectors.
Thus, we can see that $\left(\mathcal{T}d_2\right)^2 = e^{\pm 2\pi i q_2}$ and $\left(\mathcal{T}d_3 \right)^2 = e^{\pm 2\pi i q_3}$, which imply $\left(\mathcal{T}d_2\right)^2 = -1$ when $q_2 = 1/2$ and $\left(\mathcal{T}d_3 \right)^2 = -1$ when $q_3=1/2$.

Although the magnon is not a spin-$1/2$ excitation, a pair of sublattices $\{(1,2),(3,4)\}$ for $\mathcal{T}d_2$ and $\{(1,4),(2,3)\}$ for $\mathcal{T}d_3$ act as a pseudospin-1/2, $\{ \uparrow, \downarrow \}$.
Therefore $(\mathcal{T}d)^2 = -1$ results in the Kramers degeneracy which guarantees the band crossing at the magnetic glide invariant momenta with $q_{2,3} = 1/2$; we get two straight lines of two-fold degenerate momentum points,
\begin{align}
\mathbf{q} = \frac{1}{2} (\mathbf{b}_1 +\mathbf{b}_2) + q (\mathbf{b}_1 + \mathbf{b}_3)
\end{align}
for $(\mathcal{T}d_2)^2 = -1$, and
\begin{align}
\mathbf{q} = \frac{1}{2} (\mathbf{b}_1 +\mathbf{b}_3) + q (\mathbf{b}_1 + \mathbf{b}_2)
\end{align}
for $(\mathcal{T}d_3)^2 = -1$.

Note that the magnetic glide \emph{enforces} the band crossing.
This implies that this band crossing must occur even if we turn off the pairing ($\mathcal{H}_\mathrm{hop}$ only) [Fig.~\ref{fig:magnon} (c)].
After we analytically diagonalize $\mathcal{H}_\mathrm{hop}$,
\begin{align}
&E_n(\mathbf{q})= a_0 \pm \Big [ b_0^2 + x_0^2 (A_{11}^2 + A_{12}^2 + A_{21}^2 + A_{22}^2) \nonumber \\
&\pm 2 \sqrt{b_0^2 x_0^2 (A_{11}^2+A_{21}^2)+x_0^4 (A_{11}A_{22}-A_{12}A_{21})^2}  \Big ]^{\frac{1}{2}},
\label{eq:EkSU2}
\end{align}
we can see that the first and second band touching and the third and fourth band touching happen when
\begin{align}
A_{11} &= \frac{1}{2} \left(1+ \cos (2\pi q_1) + \cos (2\pi q_2) + \cos (2\pi q_3) \right) = 0, \label{eq:A11-0} \\
A_{21} &= \frac{1}{2} \left( \sin (2\pi q_1) + \sin (2\pi q_2) +\sin (2\pi q_3) \right) =0, \label{eq:A21-0}
\end{align}
which implies three one-dimensional nodal lines
\begin{align}
\mathbf{q} &= \frac{1}{2}\left(\mathbf{b}_1 + \mathbf{b}_2 \right) + q_u (\mathbf{b}_1 + \mathbf{b}_3), \label{eq:uNL}\\
\mathbf{q} &= \frac{1}{2}\left(\mathbf{b}_1 + \mathbf{b}_3 \right) + q_v (\mathbf{b}_1 + \mathbf{b}_2), \label{eq:vNL} \\
\mathbf{q} &= \frac{1}{2} \left( \mathbf{b}_2+\mathbf{b}_3 \right) + q_w (\mathbf{b}_2 + \mathbf{b}_3). \label{eq:wNL}
\end{align}
The third nodal line band touching [Eq.~(\ref{eq:wNL})] along $\mathbf{b}_2 + \mathbf{b}_3$ direction [the straight blue lines in Fig.~\ref{fig:magnon} (c)] is accidental and eventually lifted as soon as we introduce magnon pairing.
However, the first two nodal lines are robust even if we include $\mathcal{H}_\mathrm{pair}$.

\textit{Glide protected topological nodal lines.} --
There are another kind of nodal lines between the first (third) and second (fourth) bands at $q_z = 0$ plane.
Unlike the magnetic glide protected nodal lines, these nodal lines can merge into a point and vanish under sufficiently strong off-diagonal exchange (red lines in Fig.~\ref{fig:phases}).
However, it is still stable within some finite region in the parameter space because of the glide mirror $d_1$.
As the $q_z=0$ plane is invariant under $d_1$, the eigenvalue for the glide mirror is a good quantum number within the plane.
Therefore band crossing can occur between bands having different $d_1$ eigenvalues, $\varepsilon e^{- \pi i q_1}$ with $\varepsilon = \pm1$.

In spite of the benign mechanism, the presence of this nodal line can be quite puzzling if we consider the spectrum of $\mathcal{H}_\mathrm{hop}$.
Recall that $\mathcal{H}_\mathrm{hop}$ gives no band crossing at $q_z = 0$ plane other than the $\mathcal{T}d$ protected nodal lines.
In the case of fermionic superconductors, the pairing term destabilizes the Fermi surface and reduces the dimensionality of the gapless spectrum.
In order to get a gapless superconductor, the parent metallic state needs to be ``more'' gapless because the pairing introduces another channel for the mass term.
However, our most generic parent Hamiltonian, $\mathcal{H}_\mathrm{hop}$, does not have any non-trivial nodal lines (except $\mathcal{T}d$-protected nodal lines) no matter which parameters we choose.
This observation naturally raises a question 
how $d_1$ protected nodal line appears.

The solution comes from the bosonic nature of magnons.
Recall that the magnon pairing demands us to diagonalize the effective non-Hermitian matrix $\tau^3 ( \mathcal{H}_\mathrm{hop} +\mathcal{H}_\mathrm{pair})$.
Because the pairing channels for magnons introduces non-Hermitian terms which are squared to be \emph{negative} numbers, the spectrum for bosonic BdG Hamiltonian can be much richer than the bosonic Bloch Hamiltonian.
To be concrete, at $q_z=0$ plane, our effective Hamiltonian has the form
\begin{multline}
\tau^3 (\mathcal{H}_\mathrm{hop} +\mathcal{H}_\mathrm{pair})= a_0 \tau^3+ b_0 \sigma^1 \tau^3+ i b_1 \sigma^1 \tau^2 \\
+ \varepsilon \Big[ (\alpha(\mathbf{q}) \sigma^1 + \beta(\mathbf{q}) \sigma^2) (x_0\tau^3 + i x_1\tau^2) \\
+i y_2 (\gamma(\mathbf{q}) \sigma^1 + \delta (\mathbf{q}) \sigma^2 )\tau^1\Big],
\label{eq:Hkd1pair}
\end{multline}
where $\alpha, \beta, \gamma, \delta$ are momentum dependent coefficients and $\varepsilon = \pm1$ is the sign of $d_1$ eigenvalue, $\varepsilon e^{\pi i q_1}$ \cite{SM}.
If we diagonalize the non-Hermtian matrix, the energy spectrum has following structure:
\begin{multline}
E_n(\mathbf{q})= \bigg ( a_0^2 + (b_0 +\varepsilon x_0 \alpha)^2 + \varepsilon^2 x_0^2 \beta^2 \\
- \Big \{ (b_1 +\varepsilon x_1 \alpha)^2 + \varepsilon^2 x_1^2 \beta^2 +y_2^2 (\gamma^2 + \delta^2)\Big\} \\
\pm 2 \Big ( a_0^2 (b_0 + \varepsilon x_0 \alpha)^2 +\varepsilon^2 a_0^2 x_0^2 \beta^2 
+\varepsilon^2 y_2^2 \big( \varepsilon x_1 \beta \gamma - (b_1 + \varepsilon x_1 \alpha)\delta \big)^2 \\
-\Big\{ \varepsilon^2 \beta^2  \big( x_1(b_0 +\varepsilon x_0 \alpha)-x_0(b_1 +\varepsilon x_1 \alpha ) \big) ^2 \\
+\varepsilon^2 y_2^2 \big( \varepsilon x_0 \beta \gamma- (b_0 + \varepsilon x_0 \alpha) \delta \big)^2 
\Big\} \Big)^{\frac{1}{2}} \bigg )^\frac{1}{2}.
\label{eq:Ekd1pair}
\end{multline}
Note all those sqaure terms in the curly brackets multiplied to a negative sign in Eq.~(\ref{eq:Ekd1pair}) come 
from the non-Hermitian pairing terms in Eq.~(\ref{eq:Hkd1pair}).

\textit{Weyl magnons.} -- Weyl points are another topological magnons which are emergent due to the magnon pairing. Although the Weyl points can exist in three spatial dimensions without any symmetry, our generic magnon Hamiltonian does not show any point-like band touching between the first and second bands in the absence of the pairing term.

All Weyl points are located at two perpendicular planes of the Brillouin zone because of the magnetic glides $\mathcal{T}d_{2,3}$ and inversion symmetry $\mathcal{P}=d_1^{-1} d_2 d_3^{-1}$; the four Weyl points between the first and second bands (red and blue points in Fig.~\ref{fig:phases}) are located at the $\mathcal{PT}d_{3}$-invariant plane, and the Weyl points between the second and third bands (orange and green points in Fig.~\ref{fig:phases}) are located at the $\mathcal{PT}d_{2}$-invariant plane.
Because the $\mathcal{PT}d_{2,3}$ require the magnon Hamiltonian $\mathcal{H}(\mathbf{q})$ at the symmetry planes to be real up to an unitary transformation, two magnon bands (with an appropriate basis choice) can be written as
\begin{align}
f_0(\mathbf{q}) \mu^0+ f_1(\mathbf{q}) \mu^1 + f_3(\mathbf{q}) \mu^3,
\end{align}
where $\mu^\alpha$ are the Pauli matrices and $f_\alpha(\mathbf{q})$ are real-valued functions.
Since we have two variable momenta at the symmetry planes and two equations, $f_1(\mathbf{q}) = f_3(\mathbf{q}) = 0$, for the band crossing, the $\mathcal{PT}d$ symmetry can protect the zero-dimensional magnon band touching, just as the Dirac cones of the ideal graphene are protected by the $\mathcal{PT}$ symmetry.

In summary, we investigated the symmetry-protected topological magnons in the field-induced canted zig-zag order in 
the hyperhoneycomb iridate, $\beta\text{-}\mathrm{Li_2IrO_3}$. We clarify the roles of the magnetic space group symmetries
on the emergence of topological nodal lines and Weyl points in the magnon spectra. 
The generic magnon Hamiltonian is derived to show that these topological magnons require 
the pairing of magnons, which is an important consequence of the broken $SU(2)$ spin rotation symmetry of the bond-directional interactions
in Kitaev materials.
We explained how varying the strength of the interactions influences the shapes and locations of the nodal lines and Weyl points.
Since the off-diagonal exchange interactions depend on the direct overlap between $\mathrm{Ir}$ orbitals \cite{Rau_JKGmodel}, 
the evolution of topological magnons can be tested experimentally by pressurizing the sample.
It would also be interesting to study how the symmetry breaking perturbations such as the magnetic fields along $\hat{a}$- or $\hat{c}$-axis 
may gap out these gapless band touching. If the Berry curvature remains large near the small gap opening, 
the perturbations may result in a large change in the anomalous thermal Hall response.

This work was supported by the NSERC of Canada, the Canadian Institute for Advanced Research,
and the Center for Quantum Materials at the University of Toronto.
T. M. is supported by Grants-in-Aid for Scientific Research, KAKENHI, JP17H06138, MEXT, Japan.
We acknowledge Jong Yeon Lee, Kyusung Hwang, and Adarsh S. Patri for insightful discussion.

\begin{acknowledgments}
\end{acknowledgments}
\bibliography{magnon}
\clearpage
\onecolumngrid
\setcounter{equation}{0}
\setcounter{figure}{0}
\setcounter{table}{0}
\setcounter{page}{1}
\setcounter{section}{0}
\setcounter{subsection}{0}

\pagebreak
\widetext
\begin{center}
\textbf{\large Supplemental Materials for ``Symmetry-protected topological magnons in three dimensional Kitaev materials"}
\end{center}
\setcounter{equation}{0}
\setcounter{figure}{0}
\setcounter{table}{0}
\setcounter{page}{1}
\makeatletter
\renewcommand{\theequation}{S\arabic{equation}}
\renewcommand{\thefigure}{S\arabic{figure}}

\section{Estimation of the coupling strength} \label{app:estimation}

Microscopically, the coupling constants $J$, $K$, and $\Gamma$ are derived from the strong coupling expansion of the multiorbital Hubbard model \cite{JKmechanism, Rau_JKGmodel,Lee_hyperhoneycomb}.
Because of the edge-sharing octahedron oxygen cages, direct and indirect hopping integrals between $t_{2g}$ orbitals (e.g., for $z$-links) of Ir ions can be organized as
\begin{align}
h_{jk}^z = \hat{d}_j^\dagger
\begin{pmatrix}
t_1 & t_2 & 0 \\
t_2 & t_1 & 0 \\
0 & 0 & t_3
\end{pmatrix}
\hat{d}_k,
\end{align}
where $\hat{d}_j = (d_{j,yz}, d_{j,zx}, d_{j,xy})^T$ (spin index is suppressed).
Then the coupling strengths are given by the following formulae \cite{Rau_JKGmodel,Lee_hyperhoneycomb}:
\begin{align}
J &= \frac{4}{27} \left [ \frac{(2t_1+t_3)^2(4J_H +U)}{U^2} - \frac{16 J_H (t_1-t_3)}{(2U+3\lambda)^2} \right], \\
K &= \frac{32J_H}{9} \left[ \frac{(t_1-t_3)^2-3t_2^3}{(2U+3\lambda)^2} \right], \\
\Gamma &= \frac{64J_H}{9} \frac{t_2(t_1-t_3)}{(2U+3\lambda)^2},
\end{align}
where $U$ is the intra-orbital Coulomb repulsion, $J_H$ is the Hund coupling, and $\lambda$ quantifies the strength of spin-orbit coupling.

Based on first-principle calculations \cite{Kim_DFT}, we choose $U=3.0$ eV, $J_H = 0.2U$, and $\lambda = 0.516$ eV.
With the parameterization
\begin{equation}
(J,K,\Gamma) = J_0 (\sin \theta \cos \phi, \sin \theta \sin \phi, \cos \theta),
\end{equation}
we get $J_0 = 8.03$ meV and $(\theta, \phi) = (0.60 \pi, 1.52\pi)$, which give $K<0$, $\Gamma<0$, and $J>0$.
Because the Kitaev interaction $K$ is dominant, we take $|K|=1$ as a unit of energy. Then $(J,K,\Gamma) = (0.063, -1, -0.33)$ are the most relevant choice of parameters for $\beta$-$\mathrm{Li_2IrO_3}$.

\section{$Fddd$ space group symmetry} \label{app:Fddd}

An interacting spin Hamiltonian on the hyperhoneycomb lattice respects $Fddd$ space group symmetry, which is generated by three glide planes $d_1$, $d_2$, and $d_3$.
The glide mirror planes for $d_1$, $d_2$, and $d_3$ are all passing through the bond center of the neighboring $l=2$ and $l=3$ sites and orthogonal to the $\hat{b}$, $\hat{c}$, and $\hat{a}$-axis, respectively.
Due to spin-orbit coupling, the symmetry transformations not only transform lattice sites but also rotate local moments accordingly.

With the primitive lattice vectors $\mathbf{a}_1$, $\mathbf{a}_2$, and $\mathbf{a}_3$ [Fig.~\ref{fig:magnon} (a)], location of each spin, $\mathbf{r} = x \mathbf{a}_1 + y\mathbf{a}_2 + z\mathbf{a}_3 + \mathbf{\delta}_l $, can be labelled with integers $x, y, z \in \mathbb{Z}$ and the sublattice index $l=1,2,3,4$. Then
\begin{align}
d_1: &(x, y, z, l) \to
\begin{cases}
(x+y+z, -z, -y, 3), & l = 1 \\
(x+y+z, -z, -y, 4), & l = 2 \\
(x+y+z+1, -z, -y, 1), & l = 3 \\
(x+y+z+1, -z, -y, 2), & l = 4,
\end{cases} \nonumber \\
&(S^x_j, S^y_j, S^z_j) \to (-S^x_{d_1(j)},-S^y_{d_1(j)},S^z_{d_1(j)}),
\label{eq:d1}
\end{align}
\begin{align}
d_2 : &(x, y, z, l) \to
\begin{cases}
(-z,x+y+z,-x,3), & l = 1 \\
(-z,x+y+z,-x,4), & l = 2 \\
(-z,x+y+z+1,-x,1), & l = 3 \\
(-z,x+y+z+1,-x,2), & l = 4,
\end{cases} \nonumber \\
&(S_j^x, S_j^y, S_j^z) \to (-S_{d_2(j)}^y,-S_{d_2(j)}^x,-S_{d_2(j)}^z),
\label{eq:d2}
\end{align}
\begin{align}
d_3 : &(x, y, z, l) \to
\begin{cases}
(-y,-x,x+y+z,4), & l = 1 \\
(-y,-x,x+y+z,3), & l = 2 \\
(-y,-x,x+y+z+1,2), & l = 3 \\
(-y,-x,x+y+z+1,1), & l = 4.
\end{cases} \nonumber \\
&(S_j^x, S_j^y, S_j^z) \to (S_{d_3(j)}^y,S_{d_3(j)}^x,-S_{d_3(j)}^z),
\label{eq:d3}
\end{align}

With the Fourier transformation $S^\alpha_{jl} = \frac{1}{\sqrt{N}} \sum_\mathbf{q} S^\alpha_{\mathbf{q}l} e^{-i \mathbf{q} \cdot \mathbf{R}_j}$, the symmetry transformations act on the Fourier space $\mathbf{q} = q_1 \mathbf{b}_1 + q_2 \mathbf{b}_2 + q_3 \mathbf{b}_3$ (the reciprocal lattice vector $\mathbf{b}_j$ is normalized as $\mathbf{a}_j \cdot \mathbf{b}_k = 2\pi \delta_{jk}$) as following:

\begin{align}
d_1 : & \left| q_1, q_2, q_3, l \right \rangle \to
\begin{cases}
\left | q_1,q_1 - q_3, q_1 - q_2, 3 \right \rangle, & l = 1 \\
\left | q_1,q_1 - q_3, q_1 - q_2,4\right \rangle, & l = 2 \\
e^{-2\pi i q_1 }\left | q_1,q_1 - q_3, q_1 - q_2,1 \right \rangle, & l = 3 \\
e^{-2\pi i q_1} \left | q_1,q_1 - q_3, q_1 - q_2,2 \right \rangle, & l = 4.
\end{cases}
\label{eq:d1q}
\end{align}
\begin{align}
d_2 : & \left| q_1, q_2, q_3, l \right \rangle \to
\begin{cases}
\left | q_2-q_3,q_2, q_2 - q_1, 3 \right \rangle, & l = 1 \\
\left | q_2-q_3,q_2, q_2 - q_1,4\right \rangle, & l = 2 \\
e^{-2\pi i q_2 }\left | q_2-q_3,q_2, q_2 - q_1,1 \right \rangle, & l = 3 \\
e^{-2\pi i q_2} \left |q_2-q_3,q_2, q_2 - q_1,2 \right \rangle, & l = 4.
\end{cases}
\label{eq:d2q}
\end{align}
\begin{align}
d_3 : & \left| q_1, q_2, q_3, l \right \rangle \to
\begin{cases}
\left | q_3- q_2,q_3 - q_1, q_3, 4 \right \rangle, & l = 1 \\
\left | q_3- q_2,q_3 - q_1, q_3,3 \right \rangle, & l = 2 \\
e^{-2\pi i q_3 }\left | q_3- q_2,q_3 - q_1, q_3,2 \right \rangle, & l = 3 \\
e^{-2\pi i q_3} \left | q_3- q_2,q_3 - q_1, q_3,1 \right \rangle, & l = 4.
\end{cases}
\label{eq:d3q}
\end{align}

Because the glides accompany a half-translation, the lattice translations $T_{1,2,3}$ along $\mathbf{a}_{1,2,3}$ can be generated from the glide mirrors:
\begin{align}
(d_1)^2 = T_1, ~(d_2)^2 = T_2, ~(d_3)^2 = T_3.
\end{align}
The hyperhoneycomb lattice also has inversion symmetry $\mathcal{P}$ with respect to the bond center of the neighboring sites $l=2$ and $l=3$. The inversion can be also generated by the glide mirrors, $\mathcal{P} = d_1^{-1} d_2 d_3^{-1}$.
Because all spatial symmetries commute with time-reversal symmetry $\mathcal{T}$, $JK\Gamma$ model respects $Fddd \rtimes \mathcal{T}$ on the hyperhoneycomb lattice.

In the presence of a field-induced canted zig-zag order [Fig.~\ref{fig:magnon} (a)],
\begin{align}
\mathbf{S}_{j} = \begin{cases}
S\left( \sqrt{\frac{1-m^2}{2}}, -\sqrt{\frac{1-m^2}{2}},m \right), & l=1,4, \\
S\left( -\sqrt{\frac{1-m^2}{2}}, \sqrt{\frac{1-m^2}{2}},m \right), & l=2,3,
\end{cases}
\label{eq:magnetic_order}
\end{align}
where $S=1/2$ is spin length of the local moment, $m\in [-1,1]$ is a $z$-component of the magnetic moment in the global Cartesian coordinates,
the symmetry group is spontaneously reduced to $Fdd'd'$ magnetic space group generated by one glide $d_1$ and two magnetic glides $d'_2 = \mathcal{T} d_2$ and $d'_3=\mathcal{T}d_3$.

\section{Linear spin wave theory} \label{app:LST}

\subsection{Derivation of the magnon Hamiltonian} \label{app:magnonH}
To derive the magnon Hamiltonian, we first rotate the spin operator at each site such that $z$-component of the rotated spin operator $\widetilde{S}_j^z$ is parallel to the local magnetic ordering\cite{magnonDerive}.
For a generic spin model, the local spin rotation $S_j^\alpha = R_j^{\alpha\beta} \widetilde{S}_j^\beta$ gives\begin{align}
\hat{H} &= \frac{1}{2} \sum_{jk} \sum_{\alpha,\beta=x,y,z} S_j^\alpha J_{jk}^{\alpha\beta} S_k^\beta 
= \frac{1}{2} \sum_{jk} \sum_{\alpha,\beta=x,y,z} \widetilde{S}_j^{\alpha'}  (R_j^T)^{\alpha' \alpha} J_{jk}^{\alpha\beta}  R_k^{\beta\beta'} \widetilde{S}_k^{\beta'}
\equiv \frac{1}{2} \sum_{jk} \sum_{\alpha,\beta=x,y,z} \widetilde{S}_j^\alpha \widetilde{J}_{jk}^{\alpha\beta} \widetilde{S}_k^\beta \\
&= \frac{1}{2} \sum_{jk} \bigg( \sum_{\alpha,\beta=x,y} \widetilde{S}_j^\alpha \widetilde{J}_{jk}^{\alpha\beta} \widetilde{S}_k^\beta + \widetilde{S}_j^z \widetilde{J}_{jk}^{z z} \widetilde{S}_k^z 
+ \sum_{\alpha=x,y} \widetilde{S}_j^z \widetilde{J}_{jk}^{z\alpha} \widetilde{S}_k^\alpha + \widetilde{S}_j^\alpha \widetilde{J}_{jk}^{\alpha z} \widetilde{S}_k^z \bigg). \label{eq:rotH}
\end{align}
With the linearlized Holstein-Primakoff transformation \cite{Holstein_Primakoff},
\begin{align}
\begin{pmatrix}
\widetilde{S}_j^x \\[4 pt]
\widetilde{S}_j^y
\end{pmatrix}
\approx \sqrt{\frac{S}{2}}
\begin{pmatrix}
1 & 1 \\[4 pt]
-i & i
\end{pmatrix}
\begin{pmatrix}
b_j \\[4 pt]
b_j^\dagger
\end{pmatrix},
~\widetilde{S}_j^z = S - b_j^\dagger b_j,
\end{align}
we can obtain the magnon Hamiltonian by collecting the terms up to quadratic order:
\begin{align}
&\hat{H} =\frac{1}{2} \sum_{jk}
\begin{pmatrix}
\widetilde{S}_j^x & \widetilde{S}_j^y
\end{pmatrix}
\begin{pmatrix}
\widetilde{J}^{xx}_{jk} & \widetilde{J}^{xy}_{jk} \\[4pt]
\widetilde{J}^{yx}_{jk} & \widetilde{J}^{yy}_{jk}
\end{pmatrix}
\begin{pmatrix}
\widetilde{S}_k^x \\[4pt]
\widetilde{S}_k^y
\end{pmatrix}
+\frac{S^2}{2} \sum_{jk} \widetilde{J}^{zz}_{jk} -\frac{S}{2}\sum_{jk}\widetilde{J}^{zz}_{jk} (b_j^\dagger b_j + b_k^\dagger b_k) \\
&=\frac{S}{4} \sum_{jk}
\begin{pmatrix}
b_j^\dagger & b_j
\end{pmatrix}
\left[
\begin{pmatrix}
1 & i \\[4pt]
1 & -i
\end{pmatrix}
\begin{pmatrix}
\widetilde{J}^{xx}_{jk} & \widetilde{J}^{xy}_{jk} \\[4pt]
\widetilde{J}^{yx}_{jk} & \widetilde{J}^{yy}_{jk}
\end{pmatrix}
\begin{pmatrix}
1 & 1 \\[4pt]
-i & i
\end{pmatrix}
\right]
\begin{pmatrix}
b_k \\[4pt]
b_k^\dagger
\end{pmatrix}
+E_0 -\frac{S}{2}\sum_{jk}\widetilde{J}^{zz}_{jk} (b_j^\dagger b_j + b_k^\dagger b_k) \\
&\equiv E_0 + \frac{S}{2} \sum_{jk}
\begin{pmatrix}
b_j^\dagger & b_j
\end{pmatrix}
\begin{pmatrix}
A_{jk} & B_{jk} \\[4pt]
B_{jk}^* & A_{jk}^*
\end{pmatrix}
\begin{pmatrix}
b_k \\[4pt]
b_k^\dagger
\end{pmatrix}, \label{eq:HBdGlattice}
\end{align}
where $E_0 =\frac{S^2}{2} \sum_{jk} J_{jk}^{zz}$ is the classical ground state energy of the magnetic order, and
\begin{align}
A_{jk} &=\frac{1}{2}\left[
(\widetilde{J}^{xx}_{jk}+\widetilde{J}^{yy}_{jk}) - i(\widetilde{J}^{xy}_{jk} - \widetilde{J}^{yx}_{jk})\right] - \widetilde{J}_{jk}^{zz}\delta_{jk} \\
B_{jk} &= \frac{1}{2} \left[(\widetilde{J}^{xx}_{jk}-\widetilde{J}^{yy}_{jk}) + i(\widetilde{J}^{xy}_{jk} + \widetilde{J}^{yx}_{jk}) \right].
\end{align}
The last cross terms between $\widetilde{S}^{x,y}$ and $\widetilde{S}^z$ in Eq.~(\ref{eq:rotH}) gives linear boson terms,
but they must be vanishing after we sum over all lattice sites $j,k$ in order to have stable magnetic order.
If the magnon Hamiltonian has non-vanishing linear terms, then the ground state has non-vanishing expectation value $\langle b_j \rangle$, which implies condensation of magnon and instability of the assumed magnetic order.

\subsection{Diagonalization of bosonic Bogoliubov-de Gennes (BdG) Hamiltonian}
To diagonalize the bosonic BdG Hamiltonian, we need to find a linear transformation $T(\mathbf{q})$ such that
\begin{align}
\hat{H} &= \frac{1}{2} \sum_\mathbf{q}
\begin{pmatrix}
b_\mathbf{q}^\dagger & b_{-\mathbf{q}}
\end{pmatrix}
\mathcal{H}(\mathbf{q})
\begin{pmatrix}
b_\mathbf{q} \\[4pt] b_{-\mathbf{q}}^\dagger
\end{pmatrix} = \frac{1}{2} \sum_\mathbf{q}
\begin{pmatrix}
\gamma_\mathbf{q}^\dagger & \gamma_{-\mathbf{q}}
\end{pmatrix}
T(\mathbf{q})^\dagger \mathcal{H}(\mathbf{q}) T(\mathbf{q})
\begin{pmatrix}
\gamma_\mathbf{q} \\[4pt] \gamma_{-\mathbf{q}}^\dagger
\end{pmatrix} \\
&=\frac{1}{2} \sum_\mathbf{q}
\begin{pmatrix}
\gamma_\mathbf{q}^\dagger & \gamma_{-\mathbf{q}}
\end{pmatrix}
\begin{pmatrix}
E(\mathbf{q}) & 0 \\[4pt]
0 & E({-\mathbf{q}})
\end{pmatrix}
\begin{pmatrix}
\gamma_\mathbf{q} \\[4pt] \gamma_{-\mathbf{q}}^\dagger
\end{pmatrix}
\equiv\frac{1}{2} \sum_\mathbf{q}
\begin{pmatrix}
\gamma_\mathbf{q}^\dagger & \gamma_{-\mathbf{q}}
\end{pmatrix}
\mathcal{E}(\mathbf{q})
\begin{pmatrix}
\gamma_\mathbf{q} \\[4pt] \gamma_{-\mathbf{q}}^\dagger
\end{pmatrix}
= \sum_\mathbf{q} \left(E(\mathbf{q}) + \frac{1}{2} \right)\gamma_\mathbf{q}^\dagger \gamma_\mathbf{q}
\end{align}

If $\mathcal{H}(\mathbf{q})$ is Hermitian and positive-definite, we can factorize the matrix into product of the lower-triangular matrix by the Cholesky decomposition: $\mathcal{H}(\mathbf{q}) = K(\mathbf{q})^\dagger K(\mathbf{q})$.
Because $\mathcal{E}(\mathbf{q})$ is diagonal, we can define $\sqrt{\mathcal{E}(\mathbf{q})}$. Then
\begin{align}
\mathcal{E}(\mathbf{q}) &= \sqrt{\mathcal{E}(\mathbf{q})} U(\mathbf{q})^\dagger [(K(\mathbf{q})^\dagger)^{-1} K(\mathbf{q})^\dagger][ K(\mathbf{q}) K(\mathbf{q})^{-1} ]U(\mathbf{q}) \sqrt{\mathcal{E}(\mathbf{q})} \\
&= \left( K(\mathbf{q})^{-1} U(\mathbf{q}) \sqrt{\mathcal{E}(\mathbf{q})} \right)^\dagger \mathcal{H}(\mathbf{q}) \left( K(\mathbf{q})^{-1} U(\mathbf{q}) \sqrt{\mathcal{E}(\mathbf{q})} \right)
\equiv T(\mathbf{q})^\dagger \mathcal{H}(\mathbf{q})T(\mathbf{q}).
\end{align}
While the above expression is true for any unitary transformation $U(\mathbf{q})$, not every unitary transformation results in canonical transformation, i.e., we also demand that the linear transformation $T(\mathbf{q})$ preserves the bosonic commutation relation $[ \gamma(\mathbf{q}),\gamma(\mathbf{q'})] = \delta(\mathbf{q} - \mathbf{q'})$.
To preserve the commutation relationship, $T(\mathbf{q})$ must satisfy the paraunitary condition \cite{bosonDiag}
\begin{align}
T(\mathbf{q})^\dagger \tau^3 T(\mathbf{q}) = T(\mathbf{q}) \tau^3 T(\mathbf{q})^\dagger = \tau^3.
\end{align}
From this condition we determine the unitary transformation $U(\mathbf{q})$,
\begin{align}
T(\mathbf{q})^\dagger \tau^3 T(\mathbf{q}) &= \sqrt{\mathcal{E}(\mathbf{q})}U(\mathbf{q})^\dagger (K(\mathbf{q})^{-1})^\dagger \tau^3 K(\mathbf{q})^{-1} U(\mathbf{q}) \sqrt{\mathcal{E}(\mathbf{q})} \\
&=\sqrt{\mathcal{E}(\mathbf{q})}U(\mathbf{q})^\dagger \left ( K(\mathbf{q}) \tau^3 K(\mathbf{q})^\dagger \right)^{-1} U(\mathbf{q}) \sqrt{\mathcal{E}(\mathbf{q})}.
\label{eq:paraunitarycondition}
\end{align}
If $U(\mathbf{q})$ is a unitary transformation diagonalizing $K(\mathbf{q}) \tau^3 K(\mathbf{q})^\dagger$,
\begin{align}
U(\mathbf{q})^\dagger \left(K(\mathbf{q}) \tau^3 K(\mathbf{q})^\dagger \right) U(\mathbf{q}) = L(\mathbf{q}),
\end{align}
then Eq.~(\ref{eq:paraunitarycondition}) gives $\mathcal{E}(\mathbf{q}) = \tau^3 L(\mathbf{q}).$

Note that the paraunitary condition implies
\begin{align}
T(\mathbf{q})^\dagger \mathcal{H}(\mathbf{q}) T(\mathbf{q}) = \mathcal{E}(\mathbf{q})
\Rightarrow  \left[\tau^3\mathcal{H}(\mathbf{q}) \right]T(\mathbf{q}) = \tau^3\left(T(\mathbf{q})^\dagger \right)^{-1} \mathcal{E}(\mathbf{q}) = T(\mathbf{q})
\begin{pmatrix}
E(\mathbf{q}) & 0 \\
0 & -E(-\mathbf{q})
\end{pmatrix}.
\label{eq:gEVP}
\end{align}
Therefore the spectrum of the bosonic BdG Hamiltonian is a set of eigenvalues of non-Hermitian matrix $\tau^3 \mathcal{H}(\mathbf{q})$.

\section{Symmetry constraints on the magnon Hamiltonian} \label{app:constraint}

In this section, we derive the generic magnon Hamiltonian based on the $Fdd'd'$ magnetic space group symmetry. Because we are interested in $JK\Gamma h$ model, we only consider the nearest-neighbor interactions and on-site terms.

Let's start from the following general nearest-neighbor magnon Hamiltonian:
\begin{align}
H &= S \sum_j \phi_{j1}^\dagger \cdot B \cdot \phi_{j2} + \phi_{j2}^\dagger \cdot R \cdot \phi_{j3} + \phi_{j+\hat{3},2}^\dagger \cdot G \cdot \phi_{j3} + \phi_{j3}^\dagger \cdot \bar{B} \cdot \phi_{j4} + \phi_{j+\hat{1},1}^\dagger \cdot \bar{R} \cdot \phi_{j4} + \phi_{j+\hat{2},1}^\dagger \cdot \bar{G} \cdot \phi_{j4} \nonumber \\
&-\frac{S}{2}\sum_j \mu_1 b_{j1}^\dagger b_{j1} + \mu_2 b_{j2}^\dagger b_{j2} + \mu_3 b_{j3}^\dagger b_{j3} + \mu_4 b_{j4}^\dagger b_{j4}, \\
&=\frac{S}{2} \sum_\mathbf{q} \phi_{\mathbf{q}1}^\dagger \cdot B \cdot \phi_{\mathbf{q}2} + \phi_{\mathbf{q}2}^\dagger \cdot \left(R + G e^{2\pi i q_3} \right) \cdot \phi_{\mathbf{q}3} + \phi_{\mathbf{q}3}^\dagger \cdot \bar{B} \cdot \phi_{\mathbf{q}4} + \phi_{\mathbf{q}1}^\dagger \cdot \left( \bar{R} e^{2\pi i q_1} + \bar{G} e^{2\pi i q_2} \right) \cdot \phi_{\mathbf{q}4} +\mathrm{h.c.} \nonumber \\
&-\frac{S}{2}\sum_\mathbf{q} \mu_1 b_{\mathbf{q}1}^\dagger b_{\mathbf{q}1} + \mu_2 b_{\mathbf{q}2}^\dagger b_{\mathbf{q}2} + \mu_3 b_{\mathbf{q}3}^\dagger b_{\mathbf{q}3}  + \mu_4 b_{\mathbf{q}4}^\dagger b_{\mathbf{q}4},
\label{eq:genericH}
\end{align}
where $j=(x,y,z)$ is the unit cell index, $\phi_{jl} = ( b_{jl}, b_{jl}^\dagger )^T$, and the Fourier transformation is defined as
\begin{align}
\phi_{jl} = \frac{1}{\sqrt{N}}\sum_\mathbf{q} \phi_{\mathbf{q}l} ~e^{-i \mathbf{q}\cdot \mathbf{R}_j}.
\end{align}
For clarity, we write $\phi_{\mathbf{q}l} = \phi(q_1,q_2,q_3,l)$ with $\mathbf{q} = q_1 \mathbf{b}_1 + q_2 \mathbf{b}_2 + q_3 \mathbf{b}_3$ and $\phi_{jl} = \phi(x,y,z,l)$ with $\mathbf{R}_j = x \mathbf{a}_1 + y \mathbf{a}_2 + z \mathbf{a}_3$ and $\mathbf{a}_\mu \cdot \mathbf{b}_\nu = 2\pi \delta_{\mu\nu}$.

\subsection{Constraints due to glide mirror $d_1$}
Using Eq.~(\ref{eq:d1q}),
\begin{align}
\hat{d}_1 \phi_{\mathbf{q}l} \hat{d}_1^{-1} &= \frac{1}{\sqrt{N}} \sum_j \phi_{d_1(j,l)} e^{i \mathbf{q} \cdot \mathbf{R}_j}
=
\begin{cases}
\phi(q_1, q_1-q_3, q_1-q_2, 3), & l=1\\
\phi(q_1, q_1-q_3,q_1-q_2,4), & l=2 \\
e^{-2\pi i q_1}\phi(q_1, q_1-q_3,q_1-q_2,1), & l=3 \\
e^{-2\pi i q_1}\phi(q_1, q_1-q_3,q_1-q_2,2), & l=4.
\end{cases}
\end{align}
With $\mathbf{\widetilde{q}} \equiv ( q_1, q_1 - q_3, q_1 - q_2)$, we can rewrite the Hamiltonian under the $d_1$ transformation:
\begin{align}
\hat{d}_1 H \hat{d}_1^{-1} &= \frac{S}{2} \sum_\mathbf{q} \phi_{\mathbf{\widetilde{q}},3}^\dagger \cdot B \cdot \phi_{\mathbf{\widetilde{q}},4} +  \phi_{\mathbf{\widetilde{q}},4}^\dagger \cdot (R + G e^{2\pi i q_3}) e^{-2\pi i q_1} \cdot \phi_{\mathbf{\widetilde{q}},1} +  \phi_{\mathbf{\widetilde{q}},1}^\dagger \cdot \bar{B} \cdot  \phi_{\mathbf{\widetilde{q}},2} \nonumber \\
&+ \phi_{\mathbf{\widetilde{q}},3}^\dagger \cdot (\bar{R}e^{2\pi i q_1}+ \bar{G} e^{2\pi i q_2} ) e^{-2\pi i q_1 } \cdot  \phi_{\mathbf{\widetilde{q}},2}
-\frac{S}{2} \sum_\mathbf{q} \mu_1  b_{\mathbf{\widetilde{q}},3}^\dagger b_{\mathbf{\widetilde{q}},3} + \mu_2 b_{\mathbf{\widetilde{q}},4}^\dagger b_{\mathbf{\widetilde{q}},4} + \mu_3 b_{\mathbf{\widetilde{q}},1}^\dagger b_{\mathbf{\widetilde{q}},1} + \mu_4 b_{\mathbf{\widetilde{q}},2}^\dagger b_{\mathbf{\widetilde{q}},2}
\end{align}
Because the Hamiltonian invariant under $d_1$,
\begin{align}
B = \bar{B},~R=\bar{R}^\dagger,~G=\bar{G}^\dagger, ~\mu_1 = \mu_3,~\mu_2=\mu_4.
\end{align}
\subsection{Constraints due to magnetic glide $\mathcal{T} d_2$}
In momentum space, $\mathcal{T}$ flips the momentum $\mathbf{q} \to -\mathbf{q}$ and complex conjugates the constants.
Hence, from Eq.~(\ref{eq:d2q}),
\begin{align}
(\mathcal{T}\hat{d}_2) \phi_{\mathbf{q}l} (\mathcal{T}\hat{d}_2)^{-1} &=
\begin{cases}
\phi(q_3-q_2, -q_2, q_1-q_2, 3), & l=1\\
\phi(q_3-q_2, -q_2, q_1-q_2,4), & l=2 \\
e^{2\pi i q_2}\phi(q_3-q_2, -q_2, q_1-q_2,1), & l=3 \\
e^{2\pi i q_2}\phi(q_3-q_2, -q_2, q_1-q_2,2), & l=4.
\end{cases}
\end{align}
Then with $\mathbf{\widetilde{q}} = (q_3-q_2,-q_2,q_1-q_2)$,
\begin{align}
&(\mathcal{T}d_2) H(\mathcal{T}d_2)^{-1} = \frac{S}{2} \sum_\mathbf{q} \phi_{\mathbf{\widetilde{q}},3}^\dagger \cdot B^* \cdot \phi_{\mathbf{\widetilde{q}},4} +  \phi_{\mathbf{\widetilde{q}},4}^\dagger \cdot (R^* e^{-2\pi i \widetilde{q}_2} + G^* e^{-2\pi i \widetilde{q}_1}) \cdot \phi_{\mathbf{\widetilde{q}},1} +  \phi_{\mathbf{\widetilde{q}},1}^\dagger \cdot \bar{B}^* \cdot  \phi_{\mathbf{\widetilde{q}},2} \nonumber \\
&+ \phi_{\mathbf{\widetilde{q}},3}^\dagger \cdot (\bar{R}^*e^{-2\pi i \widetilde{q}_3}+ \bar{G}^* ) \cdot  \phi_{\mathbf{\widetilde{q}},2}
-\frac{S}{2} \sum_\mathbf{q} \mu_1^*  b_{\mathbf{\widetilde{q}},3}^\dagger b_{\mathbf{\widetilde{q}},3} + \mu_2^* b_{\mathbf{\widetilde{q}},4}^\dagger b_{\mathbf{\widetilde{q}},4} + \mu_3^* b_{\mathbf{\widetilde{q}},1}^\dagger b_{\mathbf{\widetilde{q}},1} + \mu_4^* b_{\mathbf{\widetilde{q}},2}^\dagger b_{\mathbf{\widetilde{q}},2}.
\end{align}
In order to be invariant under $\mathcal{T}d_2$,
\begin{align}
\bar{B} = B^*,~R=\bar{G}^T,~G=\bar{R}^T,~\mu_1=\mu_3,~\mu_2=\mu_4.
\end{align}
\subsection{Constraints due to magnetic glide $\mathcal{T} d_3$}
The magnetic glide $\mathcal{T} d_3$ acts on the momentum space as
\begin{align}
(\mathcal{T}\hat{d}_3) \phi_{\mathbf{q}l} (\mathcal{T}\hat{d}_3)^{-1} &=
\begin{cases}
\phi(q_2-q_3, q_1-q_3, -q_3, 4), & l=1\\
\phi(q_2-q_3, q_1-q_3, -q_3,3), & l=2 \\
e^{2\pi i q_3}\phi(q_2-q_3, q_1-q_3, -q_3,2), & l=3 \\
e^{2\pi i q_3}\phi(q_2-q_3, q_1-q_3, -q_3,1), & l=4.
\end{cases}
\end{align}
After we relabel the momentum $\mathbf{\widetilde{q}} = (q_2-q_3, q_1-q_3, -q_3)$,
\begin{align}
&(\mathcal{T}d_3) H(\mathcal{T}d_3)^{-1} = \frac{S}{2} \sum_\mathbf{q} \phi_{\mathbf{\widetilde{q}},4}^\dagger \cdot B^* \cdot \phi_{\mathbf{\widetilde{q}},3} +  \phi_{\mathbf{\widetilde{q}},3}^\dagger \cdot (R^* + G^* e^{2\pi i \widetilde{q}_3}) e^{-2\pi i \widetilde{q}_3} \cdot \phi_{\mathbf{\widetilde{q}},2} +  \phi_{\mathbf{\widetilde{q}},2}^\dagger \cdot \bar{B}^* \cdot  \phi_{\mathbf{\widetilde{q}},1} \nonumber \\
&+ \phi_{\mathbf{\widetilde{q}},4}^\dagger \cdot (\bar{R}^*e^{-2\pi i (\widetilde{q}_2-\widetilde{q}_3)}+ \bar{G}^* e^{-2\pi i (\widetilde{q}_1-\widetilde{q}_3} ) e^{-2\pi i \widetilde{q}_3 } \cdot  \phi_{\mathbf{\widetilde{q}},1}
-\frac{S}{2} \sum_\mathbf{q} \mu_1^*  b_{\mathbf{\widetilde{q}},4}^\dagger b_{\mathbf{\widetilde{q}},4} + \mu_2^* b_{\mathbf{\widetilde{q}},3}^\dagger b_{\mathbf{\widetilde{q}},3} + \mu_3^* b_{\mathbf{\widetilde{q}},2}^\dagger b_{\mathbf{\widetilde{q}},2} + \mu_4^* b_{\mathbf{\widetilde{q}},1}^\dagger b_{\mathbf{\widetilde{q}},1}.
\end{align}
Hence, the magnetic glide $\mathcal{T}d_3$ demands
\begin{align}
B^T = \bar{B},~R=G^T,~\bar{R}=\bar{G}^T,~\mu_1 = \mu_4,~\mu_2=\mu_3.
\end{align}
\subsection{Constraints due to ``particle-hole" symmetry}
Based on $Fdd'd'$ magnetic space group symmetry, we found that the genetic nearest neighbor magnon Hamiltonian must satisfy 
\begin{align}
B=B^* = B^T =\bar{B},~ R= R^\dagger = \bar{R} = G^T = \bar{G}^T,~\mu_1=\mu_2=\mu_3=\mu_4 \equiv a_0,
\end{align}
i.e., $B = \bar{B}$ are real symmetric matrices and $R = \bar{R} = G^T = \bar{G}^T$ are Hermitian matrices.

In addition to the symmetries of magnetic order, the BdG Hamiltonian has built-in ``particle hole" symmetry, $\tau^1 \mathcal{H}(\mathbf{q})^T \tau^1 = \mathcal{H}(-\mathbf{q})$.
This symmetry imposes further constraints on the structure of the Hamiltonian.
\begin{align}
&B= \tau^1 B^T \tau^1 = \tau^1 B \tau^1 \Rightarrow [B,\tau^1 ] = 0 \Rightarrow B = b_0 + b_1 \tau^1,\\
&R \equiv X+iY= \tau^1 R^T\tau^1 = \tau^1 X^T \tau^1 + i \tau^1 Y^T \tau^1 = \tau^1 X\tau^1 - i\tau^1 Y \tau^1 \nonumber \\
&\Rightarrow [X,\tau^1] = 0,~\{Y,\tau^1\}=0 \Rightarrow X= x_0 +x_1\tau^1,~Y=iy_2 \tau^2,
\end{align}
where we decompose the Hermitian matrix $R$ into sum of a real symmetric matrix $X$ and pure imaginary skew-symmetric matrix $iY$.
Therefore there are only six real parameters $a_0, b_0, b_1, x_0, x_1, y_2 \in \mathbb{R}$ for non-interacting magnon Hamiltonian constrained by the magnetic space group symmetries and the built-in particle hole symmetry.

\subsection{The generic nearest-neighbor magnon Hamiltonian}

This section summarizes the results we have found.
As we discussed in the main text, we can relabel four sublattice sites $l=1,2,3,4$ with two separate 2-dimensional flavor indices $s'=0,1$ and $\sigma' = 0,1$.
Then the $Fdd'd'$-symmetry constrained generic magnon Hamiltonian has the form:
\begin{align}
\mathcal{H}_\mathrm{hop} &= a_0 + b_0 \sigma^1 +x_0\left( A_{11}(\mathbf{q}) s^1 \sigma^1 + A_{21}(\mathbf{q})  s^2 \sigma^1+ A_{12}(\mathbf{q})  s^1 \sigma^2 + A_{22}(\mathbf{q})  s^2 \sigma^2 \right),\\
\mathcal{H}_\mathrm{pair} &= b_1 \sigma^1 \tau^1 +x_1\left( A_{11}(\mathbf{q})  s^1 \sigma^1 + A_{21}(\mathbf{q})  s^2 \sigma^1 + A_{12}(\mathbf{q})  s^1 \sigma^2 + A_{22}(\mathbf{q})  s^2 \sigma^2 \right) \tau^1 \nonumber
\\&+ y_2 \left( B_{11}(\mathbf{q})  s^1 \sigma^1 + B_{21}(\mathbf{q})  s^2 \sigma^1 + B_{12}(\mathbf{q})  s^1 \sigma^2 + B_{22}(\mathbf{q})  s^2 \sigma^2\right) \tau^2,
\end{align}
where
\begin{align}
A_{11}(\mathbf{q})  &= \frac{1}{2} \left( 1+ \cos (2\pi q_1)+\cos (2\pi q_2)+\cos (2\pi q_3) \right),\\
A_{12}(\mathbf{q})  &= \frac{1}{2} \left(\sin (2\pi q_3)-\sin (2\pi q_1)-\sin (2\pi q_2) \right),\\
A_{21}(\mathbf{q})  &= -\frac{1}{2} \left(\sin (2\pi q_1)+\sin (2\pi q_2)+\sin (2\pi q_3) \right),\\
A_{22}(\mathbf{q})  &= \frac{1}{2} \left( 1+ \cos (2\pi q_3)-\cos (2\pi q_1)-\cos (2\pi q_2) \right),\\
B_{11}(\mathbf{q})  &= \frac{1}{2} \left( 1+ \cos (2\pi q_1)-\cos (2\pi q_2)-\cos (2\pi q_3) \right),\\
B_{12}(\mathbf{q})  &= \frac{1}{2} \left( \sin (2\pi q_2)-\sin (2\pi q_1)-\sin (2\pi q_3) \right),\\
B_{21}(\mathbf{q})  &= -\frac{1}{2} \left( \sin (2\pi q_1)-\sin (2\pi q_2)-\sin (2\pi q_3) \right),\\
B_{22}(\mathbf{q})  &= \frac{1}{2} \left( 1+ \cos (2\pi q_2)-\cos (2\pi q_1)-\cos (2\pi q_3) \right),
\end{align}
and $a_0, b_0, b_1, x_0, x_1, y_2 \in \mathbb{R}$ are fixed constants, $\mathbf{q} = q_1 \mathbf{b}_1+q_2 \mathbf{b}_2+q_3 \mathbf{b}_3$ with $q_1, q_2, q_3 \in (-\frac{1}{2}, \frac{1}{2}]$.

Note that six real parameters are further constrained by the positive definiteness of the magnon Hamiltonian.
If we directly construct the magnon Hamiltonian from the original interacting spin Hamiltonian, the magnon Hamiltonian is always positive definite as long as the given magnetic order is stable.

\section{Analytic solutions for the magnon spectrum}
\subsection{The spectrum of $\mathcal{H}_\mathrm{hop}$}
To compute the spectrum for the purely hopping Hamiltonian, Eq.~(\ref{eq:HSU2}), we need to find the algebraic equation which the Hamiltonian $\mathcal{H}_\mathrm{hop}$ satisfies.
\begin{align}
&\mathcal{H}_\mathrm{hop} = a_0 + b_0 \sigma^1 +x_0\left( A_{11} s^1 \sigma^1 + A_{21} s^2 \sigma^1+ A_{12} s^1 \sigma^2 + A_{22} s^2 \sigma^2 \right) \label{eq:HSU2},
\end{align}
Because
\begin{align}
\left(\mathcal{H}_\mathrm{hop} - a_0 \right)^2 = b_0^2 + x_0^2 \left( A_{11}^2 + A_{21}^2 + A_{12}^2 + A_{22}^2  \right) +2b_0 x_0 (A_{11} s^1 + A_{21}s^2 )+ 2x_0^2 \left(A_{21}A_{12} - A_{11} A_{22} s^3  \right) s^3 \sigma^3, \\
\left\{\left(\mathcal{H}_\mathrm{hop} - a_0 \right)^2 - \left[ b_0^2 + x_0^2 \left( A_{11}^2 + A_{21}^2 + A_{12}^2 + A_{22}^2  \right) \right]\right\}^2 = 4b_0^2 x_0^2 (A_{11}^2 + A_{21}^2) + 4x_0^2 \left(A_{21}A_{12} - A_{11}A_{22} \right)^2,
\end{align}
we can conclude that
\begin{align}
E_n(\mathbf{q}) = a_0 \pm \sqrt{b_0^2 +x_0^2 \left( A_{11}^2 + A_{21}^2 + A_{12}^2 + A_{22}^2  \right) \pm 2 \sqrt{b_0^2 x_0^2 (A_{11}^2 + A_{21}^2)+x_0^4(A_{21}A_{12} - A_{11}A_{22})^2}}
\end{align}

\subsection{The spectrum of $\mathcal{H}_\mathrm{hop} + \mathcal{H}_\mathrm{pair}$ at glide mirror $d_1$ invariant plane ($q_z = 0$)}

We can utilize the similar strategy to get the spectrum for the generic magnon Hamiltonian at $q_z=0$ plane.
First we organize the magnon Hamiltonian $\mathcal{H}_\mathrm{hop} + \mathcal{H}_\mathrm{pair}$ so that the glide plane symmetry is manifest.
Using following trigonometric identities,
\begin{align}
\sin x + \sin y + \sin z &= 4 \sin \left( \frac{x+y}{2} \right) \sin \left(\frac{y+z}{2} \right) \sin \left( \frac{z+x}{2}\right) + \sin (x+y+z), \\
\sin x + \sin y - \sin z &= 4 \sin \left( \frac{x+y}{2} \right) \cos \left(\frac{y+z}{2} \right)\cos \left( \frac{z+x}{2}\right) -\sin (x+y+z), \\
\cos x + \cos y + \cos z &= 4 \cos \left( \frac{x+y}{2} \right)\cos \left(\frac{y+z}{2} \right) \cos\left( \frac{z+x}{2}\right) - \cos (x+y+z), \\
\cos x + \cos y - \cos z &= 4 \cos \left( \frac{x+y}{2} \right)\sin \left(\frac{y+z}{2} \right) \sin\left( \frac{z+x}{2}\right) + \cos (x+y+z),
\end{align}
the momentum dependent coefficients $A$ and $B$ at $q_z = 0$ (equivalently $q_1 = q_2 + q_3$ in the reciprocal lattice vector basis) becomes
\begin{align}
A_{11} (\mathbf{q}) &= 2 \cos \left( \pi (q_2 + q_3) \right) \cos \left( \pi q_2 \right) \cos \left( \pi q_3 \right),\\
A_{12}(\mathbf{q}) &= -2 \cos \left( \pi(q_2 + q_3) \right) \sin \left( \pi q_2 \right) \cos \left( \pi q_3 \right), \\
A_{21}(\mathbf{q}) &= -2 \sin \left( \pi(q_2 + q_3) \right) \cos \left(\pi q_2 \right) \cos \left(  \pi q_3\right),\\
A_{22}(\mathbf{q}) &= 2 \sin \left( \pi(q_2 + q_3) \right) \sin \left( \pi q_2 \right) \cos \left(  \pi q_3\right),\\
B_{11}(\mathbf{q}) &= -2 \cos \left( \pi(q_2 + q_3) \right) \sin \left( \pi q_2 \right) \sin \left(  \pi q_3\right), \\
B_{12}(\mathbf{q}) &= -2 \cos \left( \pi(q_2 + q_3) \right) \cos \left(\pi q_2 \right) \sin \left( \pi q_3\right), \\
B_{21}(\mathbf{q}) &= 2 \sin \left( \pi(q_2 + q_3) \right) \sin \left(\pi q_2 \right) \sin \left(  \pi q_3 \right), \\
B_{22}(\mathbf{q}) &= 2 \sin \left( \pi(q_2 + q_3) \right) \cos \left( \pi q_2 \right) \sin \left(  \pi q_3\right).
\end{align}
Then
\begin{align}
A_{11} s^1 \sigma^1 + A_{21} s^2 \sigma^1 &= 2 \cos \left( \pi q_2\right) \cos \left ( \pi q_3  \right) \left[
\cos \left ( \pi (q_2+q_3)  \right) s^1 - \sin \left( \pi (q_2+q_3) \right) s^2 \right] \sigma^1,\\
A_{12} s^1 \sigma^2 + A_{22} s^2 \sigma^2 &= -2 \sin \left( \pi q_2 \right) \cos \left ( \pi q_3  \right) \left[
\cos \left ( \pi (q_2+q_3)  \right) s^1 - \sin \left( \pi (q_2+q_3) \right) s^2 \right]\sigma^2,\\
B_{11} s^1 \sigma^1 + B_{21} s^2 \sigma^1 &= -2 \sin \left(\pi q_2 \right) \sin \left ( \pi q_3  \right) \left[
\cos \left ( \pi (q_2+q_3)  \right) s^1 - \sin \left( \pi (q_2+q_3) \right) s^2 \right] \sigma^1,\\
B_{12} s^1 \sigma^2 + B_{22} s^2 \sigma^2 &= -2 \cos \left( \pi q_2 \right) \sin \left ( \pi q_3  \right) \left[
\cos \left ( \pi (q_2+q_3)  \right) s^1 - \sin \left( \pi (q_2+q_3) \right) s^2 \right] \sigma^2.
\end{align}
Note that the glide mirror operator can be written as
\begin{align}
\hat{d}_1 &= e^{\pi i (q_2 + q_3)} \left[\cos \left ( \pi (q_2+q_3)  \right) s^1 - \sin \left( \pi (q_2+q_3) \right) s^2\right] \\
&\equiv e^{\pi i (q_2 + q_3)} \hat{\varepsilon}
\end{align}
at the $d_1$-invariant plane $q_1 = q_2 + q_3$.
Because the glide mirror eigenvalue is a good quantum number at this plane, we can treat $\hat{\varepsilon} = \pm 1$ as a number on this plane.
Then we can write the magnon Hamiltonian as
\begin{align}
&\tau^3\mathcal{H}  = a_0 \tau^3 + b_0 \sigma^1 \tau^3 + i b_1 \sigma^1 \tau^2 + \varepsilon \left[ \left( \alpha(\mathbf{q}) \sigma^1 + \beta(\mathbf{q} ) \sigma^2 \right) (x_0 \tau^3 + i  x_1 \tau^2) + iy_2 \left(\gamma(\mathbf{q})\sigma^1 + \delta(\mathbf{q})\sigma^2 \right) \tau^1 \right] \\
&= a_0 \tau^3 + \left( b_0 +\varepsilon x_0 \alpha(\mathbf{q}) \right) \sigma^1 \tau^3 + i \left(b_1 + \varepsilon x_1 \alpha(\mathbf{q}) \right) \sigma^1 \tau^2 + \varepsilon x_0 \beta(\mathbf{q}) \sigma^2 \tau^3 + i \varepsilon x_1 \beta(\mathbf{q}) \sigma^2 \tau^2+ i \varepsilon y_2 \left(\gamma(\mathbf{q}) \sigma^1 + \delta(\mathbf{q})\sigma^2 \right) \tau^1,
\end{align}
where
\begin{align}
\alpha(\mathbf{q}) &= 2 \cos (\pi q_2 ) \cos (\pi q_3),\\
\beta(\mathbf{q}) &= -2 \sin (\pi q_2)\cos(\pi q_3), \\
\gamma(\mathbf{q}) &=2 \sin(\pi q_2) \sin (\pi q_3),\\
\delta(\mathbf{q}) &= 2\cos(\pi q_2) \sin(\pi q_3).
\end{align}

If $\tau^3 \mathcal{H}$ satisfies an algebraic equation $P(x) = 0$, then Eq.~(\ref{eq:gEVP}) gives
\begin{align}
P\left(\tau^3 \mathcal{H} \right) =T(\mathbf{q}) P \left( \tau^3 \mathcal{E}(\mathbf{q}) \right) T(\mathbf{q})^{-1} = T(\mathbf{q})
\begin{pmatrix}
P(E(\mathbf{q})) & 0 \\
0 & P(-E(\mathbf{-q}))
\end{pmatrix}
T(\mathbf{q})^{-1} = 0,
\end{align}
which implies $P(E(\mathbf{q})) = 0$.
Therefore if we find the equation $P(x)$ for $\tau^3 \mathcal{H}$, the spectrum corresponds to positive roots of the algebraic equation.

Since
\begin{align}
\left(\tau^3 \mathcal{H} \right)^2 &=  E_0^2 + 2a_0 (b_0 + \varepsilon x_0 \alpha(\mathbf{q}))\sigma^1 + 2 a_0 \varepsilon x_0 \beta(\mathbf{q}) \sigma^2  + 2 i \varepsilon \beta(\mathbf{q}) \left[ x_1(b_0 +\varepsilon x_0 \alpha(\mathbf{q}))-x_0(b_1 +\varepsilon x_1 \alpha(\mathbf{q}) ) \right] \sigma^3 \tau^1 \nonumber \\
&+ 2 i \varepsilon y_2 \left[ \varepsilon x_0 \beta(\mathbf{q})\gamma(\mathbf{q}) - (b_0 + \varepsilon x_0 \alpha(\mathbf{q})) \delta(\mathbf{q}) \right] \sigma^3 \tau^2 + 2 \varepsilon y_2 \left[ \varepsilon x_1 \beta(\mathbf{q}) \gamma(\mathbf{q}) - (b_1 + \varepsilon x_1 \alpha(\mathbf{q}))\delta(\mathbf{q}) \right] \sigma^3 \tau^3,
\end{align}
where $E_0^2 = a_0^2 + (b_0 +\varepsilon x_0 \alpha(\mathbf{q}))^2 - (b_1 +\varepsilon x_1 \alpha(\mathbf{q}))^2 + \varepsilon^2 \left[ (x_0^2 - x_1^2)\beta(\mathbf{q})^2 - y_2^2 (\gamma(\mathbf{q})^2 + \delta(\mathbf{q})^2) \right]$,
we can read out that the spectrum should be
\begin{multline}
E_n(\mathbf{q}) = \Big (E_0^2 \pm 2 \Big [ a_0^2 \left( (b_0 + \varepsilon x_0 \alpha(\mathbf{q}))^2 +\varepsilon^2 x_0^2 \beta(\mathbf{q})^2 \right) - \varepsilon^2 \beta(\mathbf{q})^2 \left[ x_1(b_0 +\varepsilon x_0 \alpha(\mathbf{q}))-x_0(b_1 +\varepsilon x_1 \alpha(\mathbf{q}) ) \right] ^2 \\
+\varepsilon^2 y_2^2 \left \{ \left[ \varepsilon x_1 \beta(\mathbf{q}) \gamma(\mathbf{q}) - (b_1 + \varepsilon x_1 \alpha(\mathbf{q}))\delta(\mathbf{q}) \right]^2 -\left[ \varepsilon x_0 \beta(\mathbf{q})\gamma(\mathbf{q}) - (b_0 + \varepsilon x_0 \alpha(\mathbf{q})) \delta(\mathbf{q}) \right]^2  \right\}
  \Big]^{\frac{1}{2}} \Big )^\frac{1}{2}
\end{multline}

\end{document}